\documentclass[aps,prl,twocolumn,amsmath,amssymb,showpacs]{revtex4-1}
\usepackage{graphicx}
\usepackage{dcolumn}
\usepackage{bm}

\usepackage{latexsym}
\usepackage{amsthm}
\usepackage{braket}

\usepackage[top = 1. in,bottom = 1. in, left = 1 in, right=1 in]{geometry}

 \newcommand{\arXiv}[1]{\href{http://www.arXiv.org/abs/#1}{#1}}
\usepackage[colorlinks=true, linkcolor=blue, bookmarks=true]{hyperref}

\makeatletter
\renewcommand\section{\@startsection {section}{1}{\z@}%
                  {-3.5ex \@plus -1ex \@minus -.2ex}
                  {2.3ex \@plus.2ex}%
                  {\normalfont\large\bfseries}}
\renewcommand\subsection{\@startsection{subsection}{2}{\z@}%
                   {-3.25ex\@plus -1ex \@minus -.2ex}%
                   {1.5ex \@plus .2ex}%
                   {\normalfont\bfseries}}
\makeatother

\newcommand{\beq}{\begin{equation}}
\newcommand{\eeq}{\end{equation}}
\newcommand{\ber}{\begin{array}}
\newcommand{\eer}{\end{array}}
\newcommand{\del}{\partial}

\newcommand{\ssty}{\scriptstyle}

\newcommand{\ena}{\end{eqnarray}}
\newcommand{\beqa}{\begin{eqnarray}}
\newcommand{\eeqa}{\end{eqnarray}}
\newcommand{\bea}{\begin{eqnarray}}
\newcommand{\eea}{\end{eqnarray}}

\theoremstyle{remark}

\renewcommand{\Re}{\operatorname{Re}}

\begin{document}

\title{Exact lowest-Landau-level solutions for vortex precession\vspace{.7mm}\\ in Bose-Einstein condensates}
\author {Anxo Biasi,$^{1}$ Piotr Bizo\'n,$^{2,3}$ Ben Craps,$^{4}$ Oleg Evnin$^{3,4}$\vspace{2mm}}

\affiliation{ $^{1}$Departamento de F\'\i sica de Part\'\i culas, Universidade de Santiago de Compostela
	 and Instituto Galego de F\'\i sica de Altas Enerx\'\i as (IGFAE), Santiago de Compostela, Spain\\
 $^{2}$Institute of Physics, Jagiellonian University, Krak\'ow, Poland\\	$^{3}$Department~of~Physics,~Faculty~of~Science,~Chulalongkorn~University,~Bangkok,~Thailand\\
$^{4}$Theoretische Natuurkunde, Vrije Universiteit Brussel and
	The International Solvay Institutes, Brussels, Belgium}

\begin{abstract}

The Lowest Landau Level (LLL) equation emerges as an accurate approximation for a class of dynamical regimes of Bose-Einstein Condensates (BEC) in two-dimensional isotropic harmonic traps in the limit of weak interactions. Building on recent developments in the field of spatially confined extended Hamiltonian systems, we find a fully nonlinear solution of this equation representing periodically modulated precession of a single vortex. Motions of this type have been previously seen in numerical simulations and experiments at moderately weak coupling. Our work provides the first controlled analytic prediction for trajectories of a single vortex, suggests new targets for experiments, and opens up the prospect of finding analytic multi-vortex solutions. 

\end{abstract}

\maketitle

Since the discovery of Bose-Einstein condensates (BEC) in ultracold atomic gases, considerable experimental and theoretical work has been devoted to their properties in the presence of rotation, which leads to formation of quantized vortices (for reviews, see \cite{BDZ,cooper,fetter}). While certain stationary configurations of vortices have been a subject of semi-analytic and analytic investigations \cite{ho, LLLvortex, PK}, to the best of our knowledge results on nonlinear motions of vortices have so far involved either numerics or approximations \cite{vortth1,vortth2,vortth3,vortth4,parker,vortth5}. In this article, we show that analytic progress can be made by drawing inspiration from recent developments in the field of spatially confined extended Hamiltonian systems \cite{GG,GHT,GT,CF}, which have not thus far surfaced in the BEC literature. As a first step, we find analytic solutions describing periodically modulated precession of a single vortex. We believe that a systematic generalization of our approach will eventually be used to study multi-vortex dynamics, a question of great appeal from both phenomenological and mathematical perspective.

In situations relevant for us here, one considers Bose-Einstein condensates narrowly confined in one spatial direction, so that the dynamics is effectively two-dimensional. In this two-dimensional $xy$-plane, the condensate is placed in an isotropic harmonic potential known as the `trap.' The system is described by the Gross-Pitaevskii (GP) equation for the condensate wavefunction $\Psi(t,x,y)$
\begin{equation}
i \del_t\Psi =\frac12\left(-\del_x^2-\del_y^2+x^2+y^2\right)\Psi + g|\Psi|^2\Psi,
\label{GP}
\end{equation}
where $g$ is a dimensionless coupling constant, proportional to the atomic scattering length and the total number of atoms (we impose $\int |\Psi|^2 dx dy=1$).
Our focus will be on studying this equation in the {\it weakly nonlinear regime} $0<g\ll 1$ \cite{g}. Positions of condensate vortices are given by the zeroes of $\Psi$.

A key feature for the weakly nonlinear dynamics of (\ref{GP}) is that the eigenmodes of the linearized problem ($g=0$) oscillate with integer frequencies, and consequently arbitrarily small nonlinearities produce significant effects over long times, due to the presence of resonances. To deal with this situation, the {\it time-averaging} method \cite{murdock} is particularly suitable. One starts by going to the interaction picture, which amounts to expanding $\Psi$ in the form
\begin{equation}
\Psi(t,r,\phi) = \sum_{nm} \alpha_{nm} (t)\,e^{-iE_nt} e^{im\phi} \chi_{nm}(r),
\label{expand}
\end{equation}
where $e^{im\phi} \chi_{nm}(r)$ are normalized isotropic harmonic oscillator eigenstates of energy $E_n=n+1$ and angular momentum $m\in\{-n,-n+2,...,n-2,n\}$. Substituting (\ref{expand}) to (\ref{GP}), one gets
\begin{equation}
i\,\frac{d\alpha_{nm}}{dt}\!=\! g\hspace{-8.5mm}  \sum_{\begin{array}{c}\ssty n_1,n_2,n_3\geq 0\vspace{-1.5mm}\\\ssty m+m_1=m_2+m_3\end{array}}\hspace{-7mm}\!
C_{n n_1 n_2 n_3}^{m m_1 m_2 m_3} \bar\alpha_{n_1m_1}\alpha_{n_2m_2}\alpha_{n_3m_3} e^{-i E t},
\label{GP_beta}
\end{equation}
where  the interaction coefficients $C$ are expressible through integrals of products of the eigenfunctions $\chi_{nm}$ and  $E=E_n+E_{n_1}-E_{n_2}-E_{n_3}$. The terms with $E=0$ correspond to resonant interactions while those with $E\neq 0$ are non-resonant.
Time-averaging consists in introducing the \emph{slow time} $\tau=g t$ and dropping in \eqref{GP_beta} all non-resonant terms, which oscillate rapidly in terms of $\tau$. The resulting equation (called the time-averaged or the resonant system) takes the  form
\begin{equation}
i\,\dot\alpha_{nm}=\hspace{-7mm}\sum_{\begin{array}{c}\ssty n+n_1=n_2+n_3\vspace{-1.5mm}\\\ssty m+m_1=m_2+m_3\end{array}}\hspace{-6mm}\!
C_{n n_1 n_2 n_3}^{m m_1 m_2 m_3}\bar\alpha_{n_1m_1}\alpha_{n_2m_2}\alpha_{n_3m_3},
\label{resGP}
\end{equation}
where from here onward an overdot denotes $d/d\tau$. 

It can be proved  that for sufficiently small $g$ the resonant system \eqref{resGP}  provides an accurate approximation to the original system  within any time interval of order  $1/g$ \cite{murdock}. More specifically, for any given $T$ there exist finite $c$ and $g_1$ such that the norm of the difference between solutions to (\ref{GP_beta}) and (\ref{resGP}) starting with the same initial conditions at $t=0$ will remain uniformly smaller than $cg$ at all times $t<T/g$ for any $g<g_1$. (In more qualitative terms, any given error standard can be met by our approximation on long time intervals by lowering the coupling to a sufficiently small, finite value.) Note that this property is highly non-trivial, since resonant interactions can produce effects of order 1 on time scales of order $1/g$ for arbitrarily small $g$. The key feature of the approach taken here is that (\ref{resGP}) correctly keeps track of resonant interactions, while non-resonant interactions produce only contributions of order $g$ on time scales of order $1/g$. Some pedagogical comments on time-averaging can be found in \cite{CEV2}, and its application to spherically symmetric solutions of (\ref{GP}) can be found in \cite{BMP}. We note that the time-averaging method and studies of the resulting resonant systems is part of the standard lore in nonlinear science and PDE analysis, but to the best of our knowledge these methods have not been applied extensively thus far in the context of BEC dynamics in harmonic traps. One of our aims here is to demonstrate that such applications are fruitful. (As examples of significant applications of resonant systems in the field of PDE analysis, see \cite{KSS, KM}.)

The fact that the sum in (\ref{resGP}) is constrained by the resonance condition $n+n_1=n_2+n_3$ and angular momentum conservation $m+m_1=m_2+m_3$ guarantees that if only modes with $m=n$ are excited in the initial state, no other modes will get excited in the course of evolution. These maximally rotating modes are known as the Lowest Landau Level (LLL) modes due to analogies with motion of a charged particle in a constant magnetic field \cite{fetter}. Restricting (\ref{resGP}) to these modes results in the LLL equation \cite{GHT,GT}
\begin{equation}\label{LLL_alpha}
  i \dot \alpha_n =
  \sum\limits_{j=0}^{\infty} \sum_{k=0}^{n+j} S_{njk,n+j-k} \bar \alpha_j \alpha_k \alpha_{n+j-k},
\end{equation}
where $\alpha_n\equiv \alpha_{nn}$, and the interaction coefficients $S$ are given by
\begin{equation}
S_{njk,n+j-k}=\frac{1}{2\pi} \frac{(n+j)!}{2^{n+j} \sqrt{n!j!k!(n+j-k)!}}.
\end{equation}

We remark that projecting on LLL modes is most commonly used as a variational ansatz for the condensate ground states \cite{ho,LLLvortex}. In contrast, we are using it to discuss fully dynamical solutions of (\ref{GP}), and the approximation provided by the LLL equation (\ref{LLL_alpha}) is protected by precise mathematical results on time-averaging. More specifically, (\ref{LLL_alpha}) is a consistent truncation of (\ref{resGP}) and, starting from initial conditions containing only LLL modes, no non-LLL modes will be generated at any future times in the evolution defined by  (\ref{resGP}). Furthermore, (\ref{resGP}) approximates the full Gross-Pitaevskii equation (\ref{GP_beta}) in the precise mathematical sense we have outlined above. Hence, if one evolves initial data containing only LLL modes with the full Gross-Pitaevskii system in the weak coupling regime $g\ll 1$, it is guaranteed that the amplitudes of non-LLL modes will remain small (of order $g$) over long time scales (of order 1/g). We emphasize that this picture of LLL decoupling presents a significant improvement in terms of rigor over the usual heuristic energy-ratio estimates in the style of \cite{ho}. We note furthermore that a straightforward generalization of our arguments demonstrates consistent decoupling of any other Landau level in the weak coupling regime and, more generally, of any subset of modes in (\ref{GP_beta}) satisfying $n=cm+d$ with arbitrary $c$ and $d$. (Such decoupling, while being a consequence of mathematical theorems in our context, would be very difficult to justify by the conventional heuristics based on differences of level energies.)

The LLL equation is structurally similar to many other interesting equations arising in mathematical physics. Examples include the cubic Szeg\H o equation \cite{GG} studied as an integrable model of weak turbulence, the resonant system \cite{CEV1,CEV2} for weakly nonlinear perturbations of Anti-de Sitter spacetime \cite{BR,BMR,rev2}, or the conformal flow \cite{CF} describing weakly nonlinear solutions of the conformally coupled cubic wave equation on a 3-sphere. Our subsequent analysis of the LLL equation will display intriguing parallels to some of these systems. (We note that the Gross-Pitaevskii equation (\ref{GP}) emerges as a non-relativistic limit of wave equations in anti-de Sitter spacetime \cite{CEL}, see also \cite{BMP}. This limit underlies some of the structural parallels we have just mentioned.)

It is convenient to introduce a complex variable $z=x+iy$, and the {\it critically rotating frame}, which rotates around the origin with angular velocity 1 (in this frame, the centrifugal force exactly cancels the harmonic trapping force). The most general LLL wavefunction in this frame can be expressed through $\alpha_n$ as
\begin{equation}\label{decomposition}
 \psi=\sum\limits_{n=0}^{\infty} \alpha_n(\tau)\chi_n(z), \quad \chi_n(z)=\dfrac{z^n}{\sqrt{\pi n!}} e^{-\frac{1}{2}|z|^2}.
\end{equation}
Here, $\chi_n(z)=e^{i n\phi} \chi_{nn}(r)$, and $\psi$ is related to the lab frame wavefunction $\Psi$ by $\psi(\tau,z)=e^{it} \Psi(t, e^{it} z)$.
In terms of $\psi$,
 equation
  \eqref{LLL_alpha} reads \cite{GHT}
\begin{equation}\label{LLL}
  i \dot \psi= \Pi(|\psi|^2 \psi),
\end{equation}
where $\Pi$ is the orthogonal projector on the LLL space, given explicitly by
\begin{equation}\label{Pi}
 \left(\Pi \psi\right)(z)=\frac{1}{\pi} e^{-\frac{1}{2}|z|^2}\int_{\mathbb{R}^2}  e^{\bar z' z-\frac{1}{2}|z'|^2} \psi(z')\,  dx' dy'.
\end{equation}
The LLL equation is Hamiltonian
with
\begin{align}
 H&= \frac{1}{2}  \int_{\mathbb{R}^2} |\psi|^4 dx dy \\
&= \frac{1}{2}\,\sum\limits_{n=0}^{\infty} \sum\limits_{j=0}^{\infty} \sum\limits_{k=0}^{n+j} S_{njk,n+j-k} \bar \alpha_n \bar \alpha_j \alpha_k \alpha_{n+j-k}.\nonumber
\end{align}

In addition to the time-translation invariance, the LLL equation is invariant under
 phase rotations,  space rotations and  `magnetic translations:'
\begin{align}
& \psi(\tau,z) \rightarrow e^{i\theta} \psi(\tau,z) ,\label{gauge}\\
 &\psi(\tau,z) \rightarrow \psi(\tau, e^{i\varphi} z),\label{rotation}\\
 &\psi(\tau,z) \rightarrow \psi(\tau,z-q) e^{\frac{1}{2}(\bar q z-q \bar z)}, \label{mtranslation}
 \end{align}
 where $\theta, \varphi$ are real-valued, and $q$ is complex-valued.
Via Noether's theorem, these symmetries give rise to three conserved quantities,
particle number $N$, angular momentum $J$ and dipole moment $Z$:
 \begin{align}
& N =\int_{\mathbb{R}^2} |\psi|^2 \, dx dy= \sum\limits_{n=0}^{\infty} |\alpha_n|^2, \\
& J= \int_{\mathbb{R}^2} (|z|^2-1) |\psi|^2\, dx dy=\sum\limits_{n=0}^{\infty} n |\alpha_n|^2,\\
& Z= \int_{\mathbb{R}^2} z |\psi|^2 \, dx dy=\sum\limits_{n=0}^{\infty} \sqrt{n+1} \, \alpha_n \bar \alpha_{n+1}.
\end{align}
The LLL equation is also invariant under scaling $\psi(\tau,z) \rightarrow c \,\psi(|c|^2 \tau,z)$  but this symmetry   will play no role here because the scale is fixed by our choice of normalization $N=1$.

Note that each single mode $\chi_n(z)$ gives rise to a stationary solution of the LLL equation
\begin{equation}\label{one-mode}
  \psi(\tau,z)= \chi_n(z) e^{-i\lambda_n  \tau}, \quad \lambda_n=\frac{1}{2\pi} \frac{(2n)!}{2^{2n}(n!)^2}.
\end{equation}
Acting on these single-mode solutions with the symmetries, one gets two-parameter families
\begin{equation}\label{one-mode-orbit}
 \psi(\tau,z)= \frac{(z-q)^n}{\sqrt{\pi n!}}  e^{\bar q z-\frac{1}{2} |q|^2+i\theta} e^{-\frac{1}{2} |z|^2} e^{-i\lambda_n  \tau}.
\end{equation}
The $n=0$ Gaussian family is distinguished by the fact that it saturates Carlen's inequality \cite{C}
\begin{equation}\label{carlen}
   \int_{\mathbb{R}^2} |\psi|^4 dx dy \leq \frac{1}{2\pi} \left(\int_{\mathbb{R}^2} |\psi|^2 dx dy\right)^2,
\end{equation}
 hence it maximizes $H$ for fixed $N$. As a consequence, this Gaussian state is orbitally stable \cite{GHT}.

We now turn to non-trivial dynamical solutions of the LLL equation, which are the principal novel element in our presentation. The key observation is that the following single vortex ansatz
\begin{equation}\label{ansatz}
  \psi(\tau,z)=(b(\tau)+a(\tau)z)\,e^{p(\tau)z}\, e^{-\frac{1}{2} |z|^2},
\end{equation}
where $b(\tau)$, $a(\tau)$ and $p(\tau)$ are complex-valued functions, is consistent with the LLL equation in the sense that it is preserved by the flow. To show this,  expand \eqref{ansatz} according to (\ref{decomposition}) to get
\begin{equation}
\alpha_n = \sqrt{\frac{\pi}{n!}}\left(b+\frac{an}{p}\right)p^n.\label{ansatz-alpha}
\end{equation}
Substituting this expression in (\ref{LLL_alpha}), dividing both sides by $p^n/\sqrt{n!}$ and using the summation identities
\begin{align}
\sum_{m=0}^{M}\frac{M!}{m!(M-m)!} m^A&=(\xi\, \del_\xi)^A \,(1+\xi)^M\Big|_{\xi=1},\\
\sum_{m=0}^{\infty} \frac{\xi^m}{m!} m^A &= (\xi\, \del_\xi)^A e^\xi,
\end{align}
one reduces both sides of (\ref{LLL_alpha}) to quadratic polynomials in $n$. Equating the three coefficients of these polynomials results in three
ordinary differential equations for $p(\tau), a(\tau), b(\tau)$:
\begin{align}
 &8i \dot p \!=\! \left(a \bar b\! + \!|a|^2 p\right) e^{|p|^2},\label{eqp}\\
 &8 i \dot a \!= \! \left[2 (1\!+\!|p|^2) |a|^2\!  +\! 3 b p \bar a\! +\! 2 \bar b \bar p a \!+\!3 |b|^2\right]  a e^{|p|^2}, \\
& 8 i \dot b \!= \!  (2b\!+\!a \bar p) \left[ (2\!+\!|p|^2) |a|^2\! +\! 2 \bar a b p\! +\!a \bar b \bar p\! +\!2 |b|^2\right] e^{|p|^2}.\label{eqb}
\end{align}
These equations could have been alternatively derived by inserting the ansatz \eqref{ansatz} into equation \eqref{LLL} and evaluating the integral on the right-hand side.
Within the ansatz (\ref{ansatz}-\ref{ansatz-alpha}), the conserved quantities take the form:
\begin{align}
  & N = \pi \big[|b|^2+(1+|p|^2) |a|^2 +2 \Re(b p \bar a) \big]e^{|p|^2} \\
  & J =  \pi \big[|p|^2 |b|^2 + (1+3|p|^2+|p|^4) |a|^2 \\
 &\hspace{3cm}+ 2 (1+|p|^2) \Re(b p \bar a)\big]e^{|p|^2}, \nonumber\\
  & Z = \pi \big[\bar p(|b|^2+(2+|p|^2)|a|^2 + \bar p \bar b a)\\
 &\hspace{3cm} + (1+|p|^2) \bar a  b \big] \, e^{|p|^2}.\nonumber
\end{align}
  Instead of $H$, it is convenient to  use the quadratic conserved quantity $S= \pi |a|^2 \, e^{|p|^2}$ which is related to  the Hamiltonian by $8\pi H=2N^2-S^2$.
We note that $|Z|^2=N J-S^2$ but the phase of $Z$ is an independent  conserved quantity, hence the system (\ref{eqp}-\ref{eqb}) is  minimally superintegrable. Among the four conserved quantities, two triples $\{H,N,J\}$ and $\{H,N,Z\}$ are in involution.

Using the above conservation laws and normalization $N=1$, we rewrite the system  (\ref{eqp}-\ref{eqb}) in the form
\begin{align}
 &8 \pi i \dot p =  \bar Z - p, \label{dp}\\
   & 8\pi i \dot a =  (Z p -J+3) a, \label{da}\\
  &8 \pi i \dot b = Z a + (Z p -J+4) b. \label{db}
\end{align}
One first integrates \eqref{dp} to get
\begin{equation}\label{psol}
  p(\tau)=\bar Z + \left(p(0)-\bar Z\right) e^{i\omega \tau},\quad \omega=\frac{1}{8\pi}.
\end{equation}
If $p(0)=\bar{Z}$, then $p$ is time-independent. This occurs for initial conditions with $a(0)=0$ or $b(0)+a(0)\bar{p}(0)=0$, which correspond to the stationary solutions \eqref{one-mode-orbit} with $n=0$ or $n=1$, respectively. For initial data with $Z=0$ we have $p(\tau)=p(0) e^{i\omega \tau}$, while equations (\ref{da}-\ref{db}) decouple and the solution reads
\begin{equation}\label{Z=0}
  a(\tau) =a(0) e^{-i\lambda\tau}, \quad b(\tau) =b(0) e^{-i(\lambda+\omega)\tau},
\end{equation}
where $\lambda=(3-J)/8\pi$.
All other solutions can be obtained from this stationary  solution \cite{footnote1} by magnetic translations \cite{translations}
\begin{equation}\label{gsol}
  p\rightarrow p+\bar{q},\,\, a\to a e^{-q p -\frac{1}{2}|q|^2}, \,\, b\rightarrow (b- q a) e^{-q p -\frac{1}{2}|q|^2},
\end{equation}
or directly  solving (\ref{da}-\ref{db}) upon substituting \eqref{psol}.

 It follows from (\ref{da}-\ref{db}) that
the position of the vortex,
  $z_{0}(\tau)=-b(\tau)/a(\tau)$,
 satisfies the equation
  $8 i \pi\dot z_{0} = z_{0}-Z$,
hence
\begin{equation}\label{zsol}
  z_{0}(\tau)=Z+ c\, e^{-i \omega \tau},
\end{equation}
which represents clockwise  rotation with frequency $\omega$ along a circle of radius $|c|$ centered at  $Z$.

We have explored stability of our solutions by perturbing them away from the ansatz (\ref{ansatz}) and evolving with the LLL equation numerically. The resulting motion tracks unperturbed solutions, providing evidence for their stability.

We dwell for a moment on the physical features of the motion our solutions describe. The condensate configurations we consider contain exactly one vortex, given by the zero of (\ref{ansatz}). In the critically rotating frame, to which (\ref{ansatz}) refers, the vortex position performs slow clockwise circular motion (\ref{zsol}) with period $\sim 1/g$. The peak of the condensate density, given by the maximum of the Gaussian envelope in (\ref{ansatz}), is located at $\bar p\hspace{.3mm}(gt)$. Similarly to the vortex position, it performs a circular motion given by (\ref{psol}), and the two circles are concentric  (see Fig.~1). We note that in the absence of nonlinearities ($g=0$), the Lowest Landau Level wavefunctions obviously do not evolve at all in the critically rotating frame. It is important to keep in mind that, while the precession period is large for small $g$, any fixed number of precession periods falls within the validity domain of our approximation, as specified in the passage under (\ref{resGP}).
 \begin{figure}[h!]
  	\begin{center}
  		\hspace{-0mm}\includegraphics[width=1.1\columnwidth]{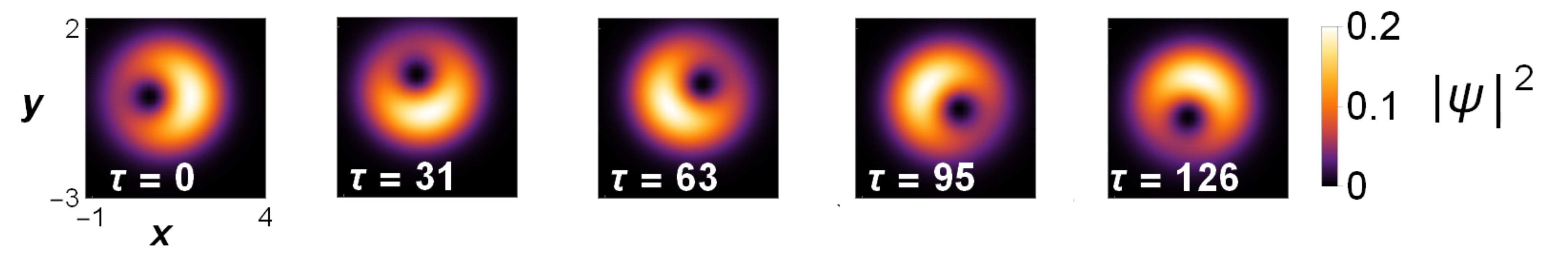}\vspace{-5mm}
  	\end{center}
  	\caption{\small{Snaphots of the condensate density $|\psi(\tau,z)|^2$ for our exact analytic solution of the LLL equation in the rotating frame. The physical time $t$ corresponds to $\tau/g$. The axis labelling is identical on all the five snapshots and only given explicitly on the leftmost one. The initial data used are $a = 0.32$, $b = -0.22$, $p = 1$.}}
  	\label{fig1}
  \end{figure}

The view in the lab frame is obtained by spinning our solutions counterclockwise around the center of the trap with angular velocity 1. In this frame, the vortex rotates around the center of the trap along a circle whose radius is slowly modulated on time scales of order $1/g$ according to \eqref{zsol}; see Fig.~2. In the special case  \eqref{Z=0} there is no modulation, only the angular velocity of the vortex is shifted away from the critical  value to $1-g/8\pi$, and the motion looks similar to Fig.~1 even if viewed from the lab frame. (This special case is reminiscent of asymmetric vortex solutions at finite coupling treated in \cite{PK}.)
    \begin{figure}[h!]
    	\begin{center}
    		\includegraphics[width=0.6\columnwidth]{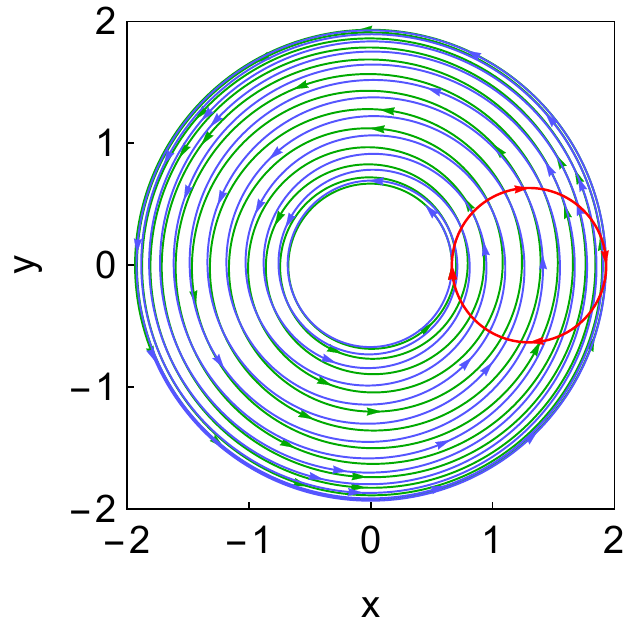}\vspace{-5mm}
    	\end{center}
    	\caption{\small{Trajectory of the vortex in the lab frame for the same initial conditions as in Fig.~1 and coupling constant $g=1$, chosen for illustrative purposes.  The radius of the orbit is slowly modulated with the frequency $g\omega\approx 0.04$. The in- and out-spiralling phases are plotted in blue and green, respectively. The red circle depicts the orbit in the rotating frame.} }
    	\label{fig2}
    \end{figure}

Vortex precession around the center of harmonic traps has been discussed in the literature on BEC experiments  \cite{vortexp1,vortexp2}, and treated with approximate analytics and numerics \cite{vortth1,vortth2,vortth3,vortth4,parker,vortth5}, at finite values of the coupling parameter $g$. The described vortex motion is a combination of circular precession around the center of the trap and jitter. This is consistent with the solutions we have derived here, if one views the jitter at finite coupling as an analog of our slow modulations at weak coupling.  In analytic treatments available in the literature, one employs approximations whose errors are not controllable, even if the results qualitatively agree with experiments, such as the Thomas-Fermi limit or matched asymptotic expansions. Our current derivations, on the other hand, while specifically tuned to the weakly nonlinear regime $g\ll 1$, are rigorous and precise. Vortex precession rates observed in the experiments and numerics are considerably below the critical angular velocity. Heuristic arguments given in \cite{vortth2} suggest that the precession rate should approach the critical rotation value as the coupling is decreased, which is consistent with our present analysis.

It is beyond our immediate goals here to analyse the prospects of experimental creation of the dynamical regime we have described, though this possibility is very tantalizing. We limit ourselves to highlighting a few obvious challenges. First of all, it is essential to create trapped condensates with very weak coupling. There appear to be systematic ways to achieve this by utilizing  Feshbach resonances \cite{feshbach}.

Another challenge is to produce initial states consistent with our wavefunction ansatz. Single vortices are nucleated in practice by spinning the trap with a certain frequency $\Omega$ and letting the condensate settle to a new spinning ground state. (The reason rotation matters is that realistic traps deviate from perfect rotational symmetry. This is relevant for discussions of vortex nucleation and production of initial states for our dynamical regime. The dynamical regime described by our ansatz, on the other hand, approximates the trap as perfectly symmetric and treats its roughness as a negligible perturbation.) For a sufficiently high $\Omega$, the first vortex nucleates, while still higher values of $\Omega$ lead to bigger arrays of vortices \cite{LLLvortex}. We also point out that our ansatz (\ref{ansatz}) is a spatial shift of a linear combination of the free particle ground state and the first excited state within the lowest Landau level. One may look for protocols generating this state by a sudden shift of the trap.

Having explored the physical interpretation of our exact LLL solutions, we briefly return to the relation with earlier uses of the LLL approximation in the literature on rotating Bose-Einstein condensates, including \cite{BDZ,cooper,fetter,ho,LLLvortex}. That work was concerned with finding ground state wavefunctions, which, in the regime in which the LLL approximation has been used, displays patterns of many vertices. The restriction to the lowest Landau level arises from the radial expansion of the condensate, which decreases the effect of interactions even for sizable values of the coupling constant $g$. In contrast, our present results deal with a very different regime in which only one vortex is present, but for which the LLL equation nevertheless provides a controlled approximation. In our setting the control is not due to expansion of a condensate -- rather, the coupling is assumed to be perturbatively small to begin with. The single-vortex configurations we describe should be thought of as highly excited states from the point of view of the Hamiltonian in the rotating frame, and, as discussed above, their realization in a lab presents a new experimental target. Extending our results to multi-vortex configurations is an important theoretical goal for future work, and we will now discuss reasons to be optimistic that it is within reach.

Our solutions have important connections to other recently explored problems of mathematical physics. The LLL equation (\ref{LLL_alpha}) is identical in terms of algebraic structure to the Fourier representation of the cubic Szeg\H o equation \cite{GG} and the conformal flow equation \cite{CF}, only the coefficients differ. Both of the latter equations admit three-dimensional invariant manifolds parametrized in a form very similar to (\ref{ansatz-alpha}), and the parallel with the conformal flow is particularly strong. The dynamics of all these invariant manifolds is characterized by periodic time dependence of the spectral localization parameter $|p|$, while the dynamical returns for the LLL equation are even stronger, with $p$ itself being exactly periodic. Our preliminary studies of stability point to some qualitative differences between the LLL equation and its cousins, but it is too early to judge how far these  differences go. The abundance of analytic results for the cubic Szeg\H o equation, which is known to be integrable, makes one hopeful that further exact solutions, beyond the single-vortex regime treated here, can be obtained for the LLL equation.

\begin{acknowledgements}
\noindent {\bf Acknowledgments:} We thank A.~Paredes, D.~Pelinovsky and K.~Sacha for comments on the manuscript. This research has been supported by the
Alexander von Humboldt Foundation, BELSPO (IAP P7/37),
CUniverse research promotion project by Chulalongkorn
University (grant reference CUAASC), FWO-Vlaanderen
(Projects G020714N and G044016N), Polish National Science
Centre Grant No. DEC-2012/06/A/ST2/00397, Ministerio de
Econom\'\i a y Competitividad de Espa\~na (Project FPA2014-52218-P) and its program ``ayudas para contratos predoctorales
para la formaci\'on de doctores 2015,'' and Vrije Universiteit
Brussel (VUB) through the Strategic Research Program ``High-Energy Physics.''
\end{acknowledgements}


\begin{thebibliography}{99}
\bibitem{g} For concreteness, we assume that $g>0$ (repulsive interactions); however, in the the weakly nonlinear regime considered here, all our results hold true for $g<0$ (attractive interactions) as well.
\bibitem{footnote1}For this solution $\protect\psi(t,z)=e^{-i \lambda \tau} \psi(0,e^{-i\omega \tau}z)$. Such stationary solutions  are sometimes called traveling waves (if $\omega\neq 0$) and standing waves (if $\omega=0$).
\end{thebibliography}
\end{document}